\documentclass[twocolumn,showpacs]{revtex4}
\usepackage[T1]{fontenc}
\usepackage{bm}
\usepackage{graphicx}
\usepackage{epstopdf}
\usepackage{amsmath}
\usepackage{color}

\begin{document}

\title{Magnetic field tuning of exciton polaritons in a semiconductor microcavity}

\author{B.~Pi\k{e}tka}
\author{D.~Zygmunt}
\author{M.~Kr\'{o}l}
\author{J.~Szczytko}
\author{J.~\L{}usakowski}
\affiliation{$^{1}$Faculty of Physics, University of Warsaw, Warsaw, Poland} 

\author{M.~R.~Molas$^{1}$}
\author{A.~A.~L.~Nicolet}
\author{M.~Potemski}
\affiliation{Laboratoire National des Champs Magn\'etiques Intenses, CNRS-UJF-UPS-INSA, Grenoble, France}

\author{P.~St\k{e}pnicki}
\author{M.~Matuszewski}
\affiliation{Institute of Physics, Polish Academy of Sciences, Warsaw, Poland}

\author{P.~Zi\k{e}ba}
\author{I.~Tralle}
\affiliation{Institute of Physics, University of Rzesz\'{o}w, Rzesz\'{o}w, Poland}

\author{F.~Morier-Genoud}
\author{B.~Deveaud}
\affiliation{Ecole Polytechnique F\'ed\'erale de Lausann, Lausanne, Switzerland}

\begin{abstract}
We detail the influence of a magnetic field on exciton-polaritons inside a semiconductor microcavity. Magnetic field can be used as a tuning parameter for exciton and photon resonances. We discuss the change of the exciton energy, the oscillator strength and redistribution of the polariton density along the dispersion curves due to the magnetically-induced detuning. We have observed that field-induced shrinkage of the exciton wave function has a direct influence not only on the exciton oscillator strength, which is observed to increase with the magnetic field, but also on the polariton linewidth. We discuss the effect of the Zeeman splitting on polaritons which magnitude changes with the exciton Hopfield coefficient and can be modelled by independent coupling of the two spin components of excitons with cavity photons.
\end{abstract}



\maketitle

\section{\label{sec:Intro}Introduction}

\begin{figure*}
	\centering
	\includegraphics[width=15cm]{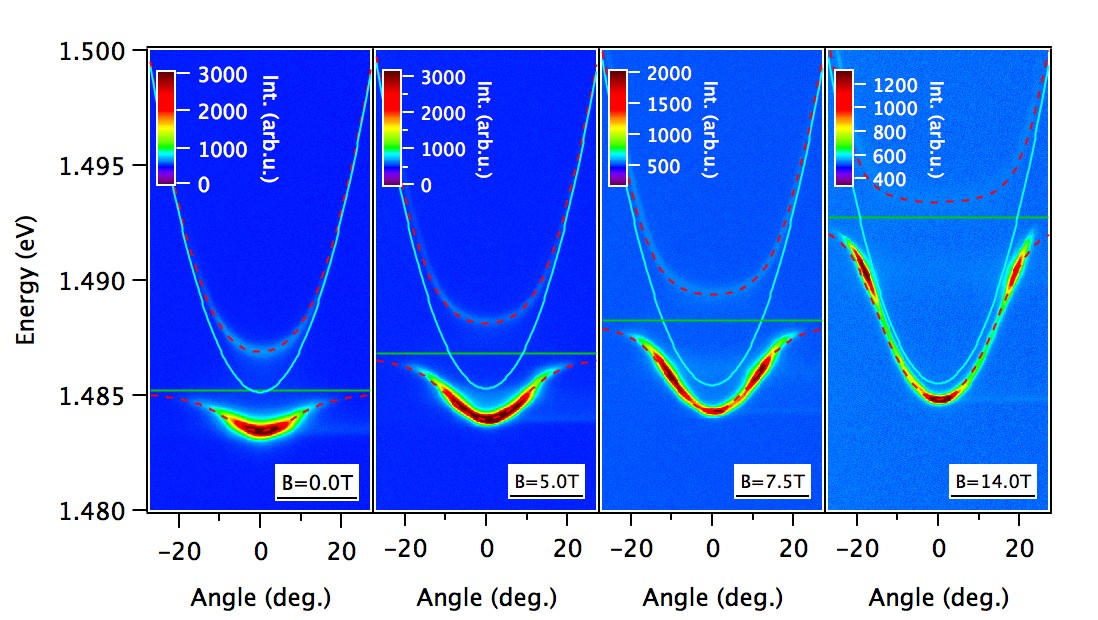} 
	\caption{Angularly resolved PL map of the exciton-polaritons at different values of magnetic field: 0~T, 5.0~T, 7.5~T and 14~T for the detuning close to zero at zero magnetic field ($\delta_{0}=$~-0.09~meV). Red dashed curves illustrate calculated LP and UP dispersions within the simple model of Rabi coupled exciton (green) and photon (white) resonances. The spectra are not resolved in polarisation. The energy splitting of polariton states is not visible due to the small value of the polariton Zeeman splitting (see Sec. \ref{sec:zeeman}).}
	\label{f1}
\end{figure*}

\begin{figure}
	\centering
	\includegraphics[width=9cm]{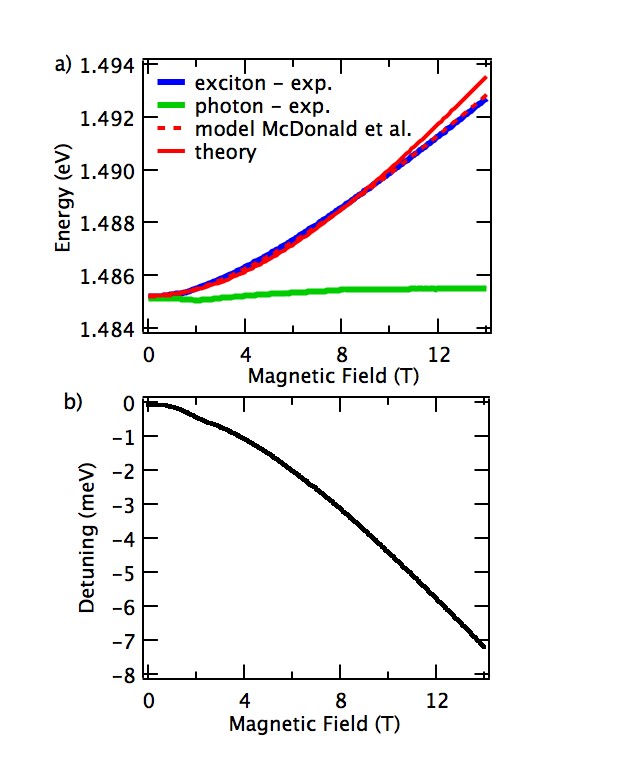}
	\caption{a) The energy change of a bare exciton (blue) and photon (green) resonances induced by magnetic field, deduced from the fit of Eq.~\ref{eq2} to the data from Fig.~\ref{f1} together with the result of the model for the exciton energy based on Ref.~\cite{MacDonald} and the calculations presented in Ref.~\cite{piotrek-Bexciton}. b) The change of photon - exciton detuning in magnetic field taken as the difference between bare photon and exciton energies from a): $\delta(B)= E_{c}-E_{x}(B)$.}
	\label{f2}
\end{figure}

\begin{figure}
	\centering
	\includegraphics[width=8cm]{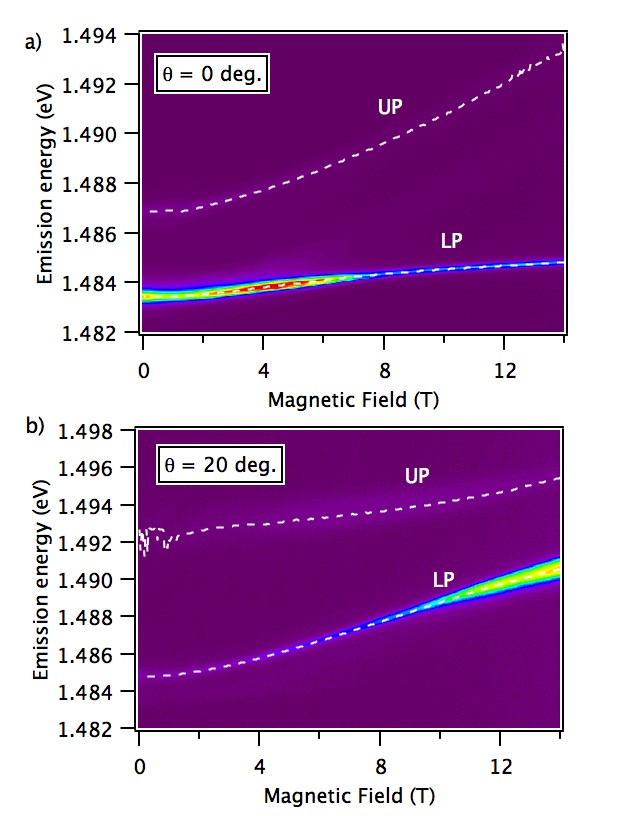}
	\caption{LP and UP polariton emission in magnetic field at a) low ($\theta$~=~0~deg.) and b) high ($\theta$~=~20~deg.) emission angles for data from Fig.~\ref{f1}. The white dashed lines correspond to the cross-sections of Fig.~\ref{f4} a) and b) for LP and UP, respectively, at corresponding angles (as marked in Fig.~\ref{f4}).}
	\label{f3}
\end{figure}

\begin{figure}
	\centering
	\includegraphics[width=8cm]{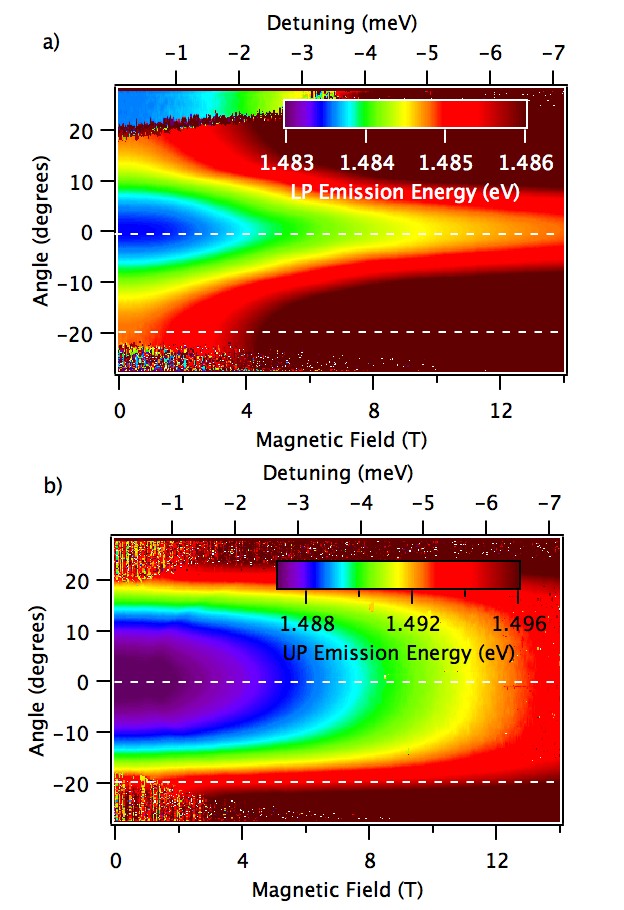}
	\caption{a) LP and b) UP emission energy for all emission angles and magnetic fields up to 14~T for data from Fig.~\ref{f1}. The cross-sections at $\theta$~=~0~deg. and $\theta$~=~20~deg. are illustrated in Fig.~\ref{f3} for LP and UP branch.}
	\label{f4}
\end{figure}

\begin{figure}
	\centering
	\includegraphics[width=9cm]{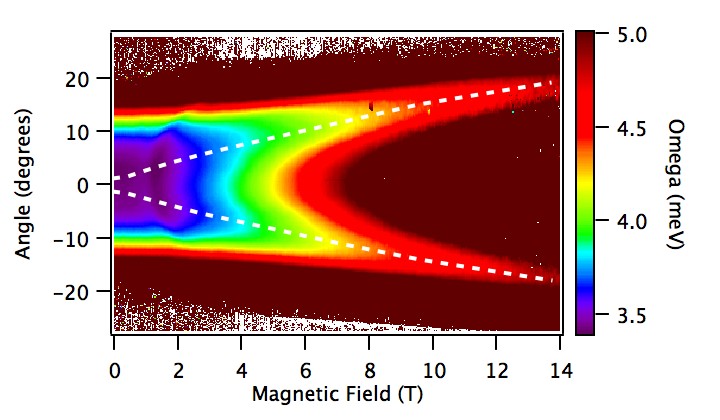}
	\caption{The difference between the UP and LP emission energies for all emission angles and magnetic field up to 14~T. The minima on the map are marked by white dotted line and correspond to the Rabi splitting.}
	\label{f5}
\end{figure}

\begin{figure}
	\centering
	\includegraphics[width=8.5cm]{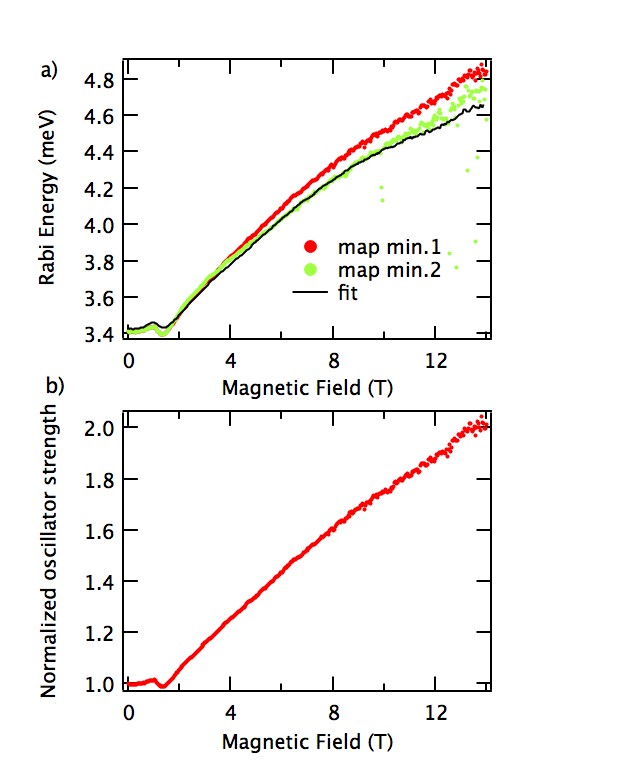}
	\caption{a) Rabi splitting obtained directly from the experiment as a minimum of the energy difference between UP and LP branches (red and green), from the fit of Eq.~\ref{eq2} to the LP and UP branches. b) Normalised exciton oscillator strength calculated from Eq.~\ref{eq4} for the experimentally determined $\Omega$ from a) (red dots).}
	\label{f6}
\end{figure}

\begin{figure*}
	\centering
	\includegraphics[width=15cm]{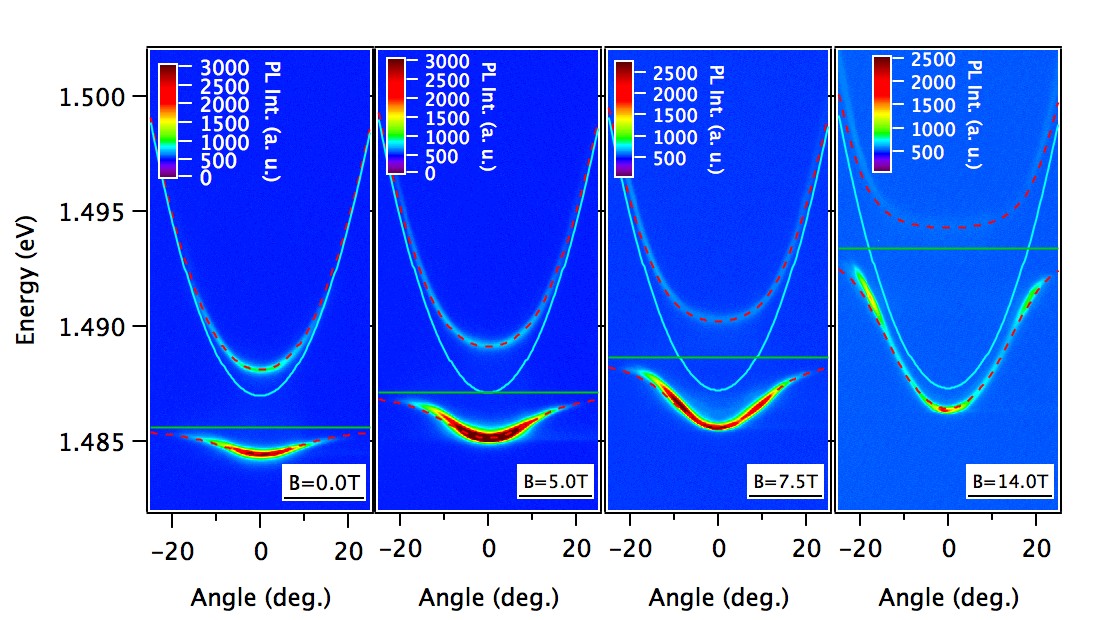} 
	\caption{Angularly resolved photoluminescence map for $\sigma+$ polarisation of the exciton-polaritons at different values of magnetic field: 0~T, 5.0~T, 7.5~T and 14~T. Similarly to Fig.~\ref{f1}, red dashed curves illustrate calculated LP and UP dispersions within the model of coupled exciton (green) and photon (white) resonances via the Rabi splitting. The Zeeman splitting is introduced in the model following to Eg.~\ref{eq5}.}
	\label{f1-1}
\end{figure*}

\begin{figure*}
	\centering
	\includegraphics[width=16cm]{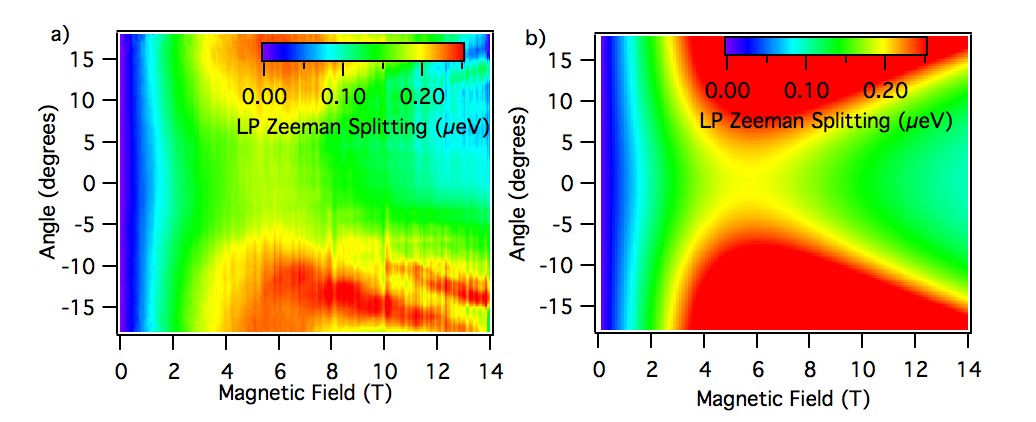}
	\caption{Polariton Zeeman splitting of LP branch for all emission angles and magnetic field up to 14T for positive exciton - photon detuning (see Fig.~\ref{f1}b). a) Experimental results  and b) model according to Eq.~\ref{eq7} .}
	\label{f9}
\end{figure*}

\begin{figure}
	\centering
	\includegraphics[width=10cm]{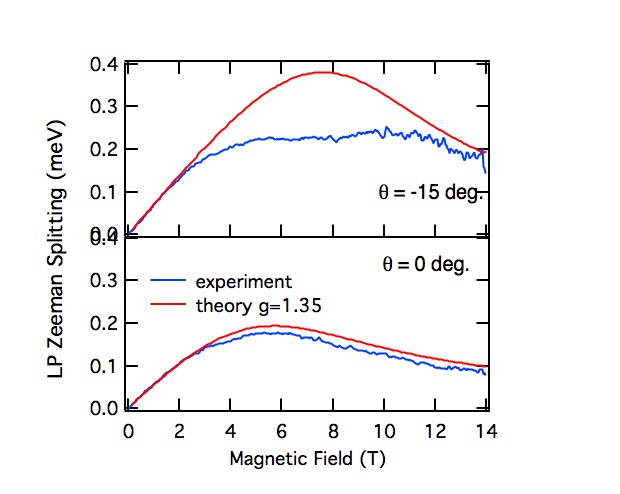}
	\caption{Comparison between experimental results (blue curve) and theoretical calculations (green curve) of LP Zeeman splitting for zero ($\theta$=~0~deg.) and high emission angle ($\theta$=~15~deg.) as marked in the figure.}
	\label{f10}
\end{figure}

\begin{figure*}
	\centering
	\includegraphics[width=16cm]{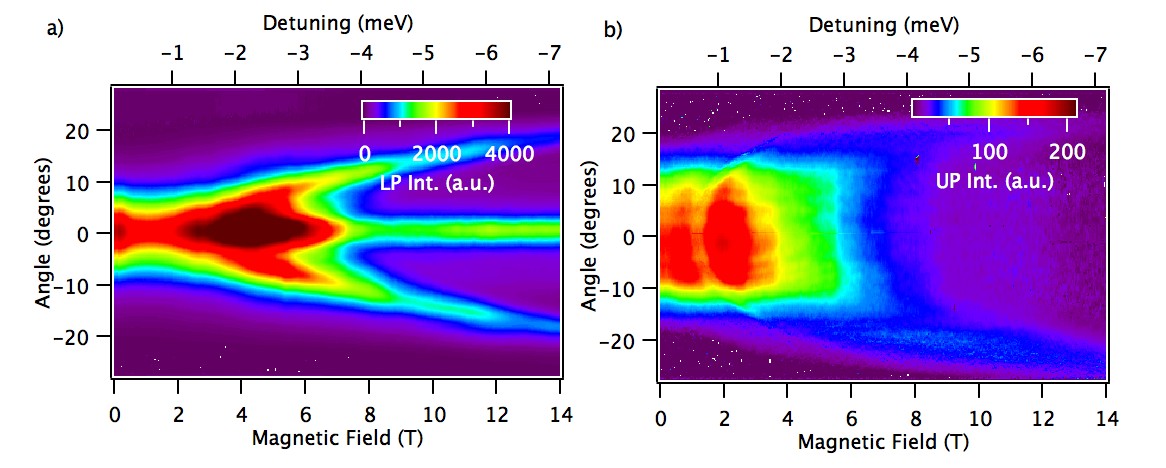}\\
	\caption{The change of a) LP and b) UP intensity in magnetic field. Eq.~\ref{eq00} describes the top axis.}
	\label{f7}
\end{figure*}

\begin{figure}
	\centering
	\includegraphics[width=9cm]{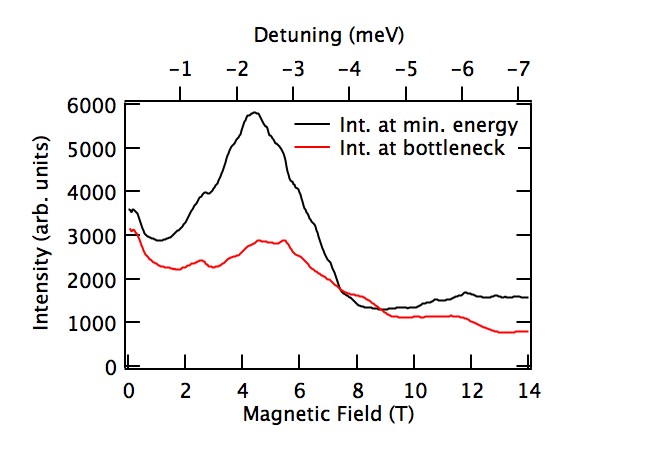}
	\caption{LP branch intensity at the bottom of the dispersion curve (minimum energy) and at the bottleneck region in magnetic field.}
	\label{f11}
\end{figure}

\begin{figure}
	\centering
	\includegraphics[width=8cm]{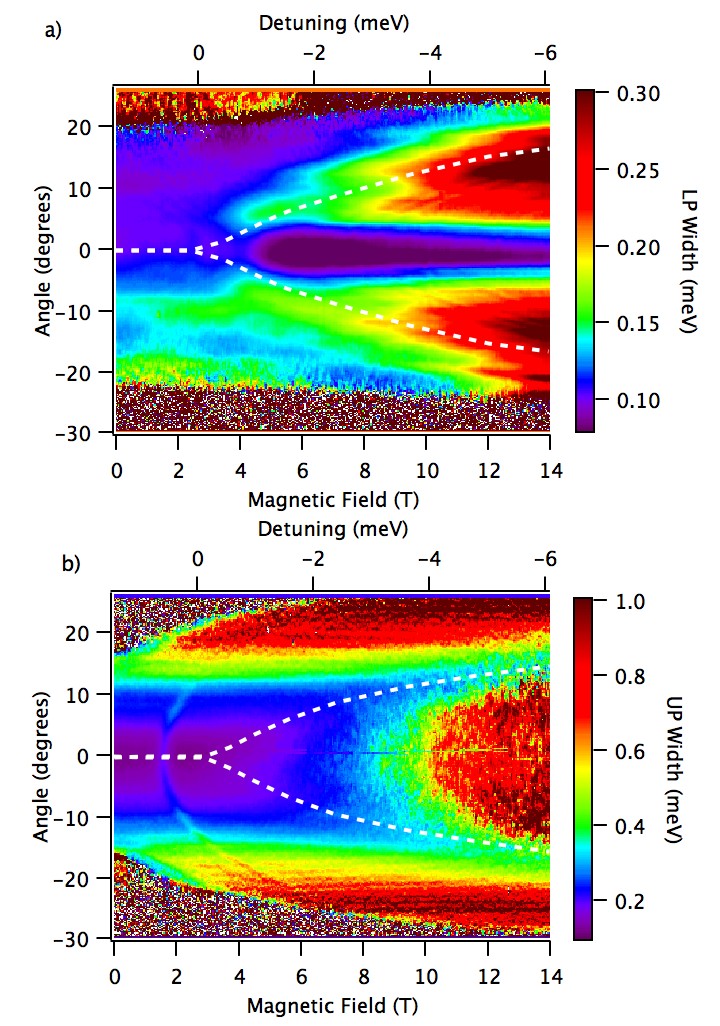}
	\caption{The change of a) LP and b) UP emission linewidth in magnetic field. The white dashed line marks the angle of zero detuning, where the Rabi splitting is determined. The top axis is given by eq.~\ref{eq00}.}
	\label{f8}
\end{figure}

\begin{figure}
	\centering
	\includegraphics[width=9cm]{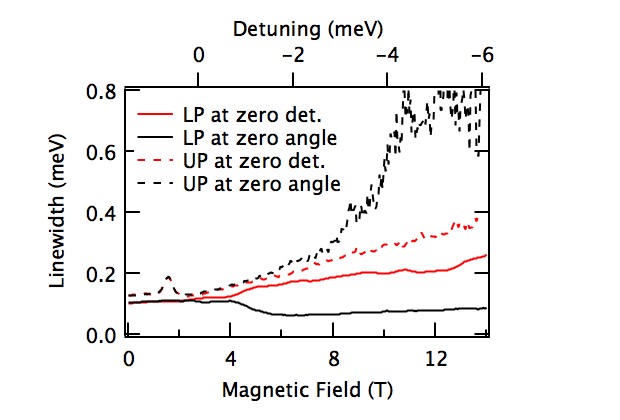}
	\caption{LP (solid line) and UP (dashed line) linewidth along at the zero detuning angle (black colour), where both branches are equally photonic and excitonic, and at the zero emission angle corresponding to the branch minimum energy (red colour) in magnetic field. The zero detuning angle is shown by the white dashed dotted line in Fig.~\ref{f8}. The increase in UP linewidth at approx. 1.5~T is due to an additional resonance from excited 2$s$HH exciton state coupled to cavity photon which perturbs the line shape of UP.}
	\label{f12}
\end{figure}

Recent interest in exciton-polariton physics has focused on quantum phenomena in these weakly interacting bosonic gases. The large number of recently demonstrated effects of exciton-polaritons in the nonlinear regime of exciton-polariton interactions include the observation of a Bose-Einstein condensate \cite{kasprzak}, the formation of a superfluid state with zero viscosity \cite{amo}, and their potential use in polariton-based devices \cite{benoit}.

Even though the field is blooming these years, reaching maturity, we notice that up to now only a few reports dealing with the impact of an external magnetic field on exciton polaritons are available \cite{whittaker, fisher, tignon, berger, armitage, forchel}. In these studies the exciton and cavity modes are tuned one with respect to the other by changing the temperature, position on the sample or electric field in order to keep the system on resonance when increasing magnetic field. In our study we do not use any additional external parameter, instead imaging the full polariton dispersion and tracing the variation of the polariton properties as a function of the magnetic field for different wavevectors. This is particularly important from the point of view of recent studies where the polariton coherent states with high in-plane momenta were induced~\cite{amo, gael, wertz}.

In this work, we focus on the effect of the magnetic field on polaritons to demonstrate that even without appreciable nonlinear effects such as bosonic stimulation, the influence of magnetic field on polaritons is diverse already in the linear regime. The magnetic field induces changes of the polariton emission energy, intensity and linewidth. The  energy of polariton states changes in magnetic field due to (i) the exciton energy shift, (ii) the modification of the exciton-photon coupling strength (the Rabi splitting, via the exciton oscillator strength) and (iii) the Zeeman splitting. The changes in the emission intensity are caused by (iv) the magnetically modified scattering process (thus thermalisation) with acoustic phonons.  In this paper we discuss the potential of magnetic field as an exciton - photon tuning parameter and address points (i) - (iv) in detail.

The paper is organised as follows. In Sec. \ref{sec:exp} we give the details of the experiment and present the sample structure. In Sec. \ref{sec:energy} we describe the effects of the polariton dispersion modification and energy shift of polariton modes induced by the magnetic field. We compare our experimental results with a theoretical model of the exciton shift in magnetic field coupled to the photonic mode. This allows us to draw conclusions about the change of the Rabi splitting in magnetic field. In Sec. \ref{sec:Rabi} we evaluate the increase of the exciton oscillator strength in magnetic field.  In Sec. \ref{sec:zeeman} we discuss the Zeeman splitting that depends on the wavevector of polaritons and changes with the excitonic content in the polariton state. Sec. \ref{sec:intensity} describes the variation of the polariton population and the effect of magnetically induced bottleneck together with the experimentally observed modification of the polariton emission linewidth.

\section{Exciton-polariton system and experimental details}
\label{sec:exp}

Exciton-polaritons are eigenmodes of strongly coupled photonic and excitonic resonances inside a semiconductor microcavity \cite{weisbuch}. To confine the photonic mode we used a GaAs lambda microcavity sandwiched between two AlAs/GaAs distributed Bragg reflectors (DBR). In the maximum of the cavity field a single 8~nm-thick In$_{0.04}$Ga$_{0.96}$As quantum well (QW) was placed, providing the excitonic component of polaritons. Excitons confined inside the QW couple to the photon modes with a coupling strength given by the Rabi splitting, $\Omega$. In the case of our sample, the excitonic resonance was at approx.~1.484~eV and the vacuum Rabi splitting was equal to approx.~$\Omega$ = 3.5~meV at zero magnetic field. The on-resonance polariton linewidth was 0.3~meV allowing for a clear resolution of polariton states. Details on the sample structure can be found in Ref.~\cite{ounsi}. The polariton population was created non-resonantly by a continuous wave external laser tuned to the first high-energy minimum of the Bragg reflectors observed in the reflectivity spectra. 

We used two different experimental setups. In the first setup, the sample was placed in a magnetic field up to 5~T in a Faraday configuration, at a cold finger of an optical cryostat at the temperature of 5~K. The photoluminescence (PL) was collected through a microscope objective of a high numerical aperture (NA=0.5). In the second setup, the sample was placed in a liquid helium bath in a superconducting cryostat up to 14~T. The emission was collected through high NA simple plano-convex lens (NA=0.63) and propagated in free space to the entrance slits of a spectrometer. In both cases, the PL was resolved angularly providing a direct measure of the dispersion of polaritons (energy vs momentum) within the same method as presented in Ref.~\cite{kasprzak}, where the emission angle, $\theta$, is proportional to the polariton in-plane momentum.

We studied the polariton emission spectra with the magnetic field up to 14~T for different exciton-phonon detunings. We define the detuning as the energy difference between the photon ($E_{c}$) and the exciton ($E_{x}$) bare resonances at zero momentum and zero magnetic field:
\begin{equation}
\delta_{0}(k_{\parallel}, B)= E_{c}(0, 0)-E_{x}(0, 0)=\delta_{0}.
\label{eq00}
\end{equation}

The first panel in Fig.~\ref{f1} illustrates an image of the angularly resolved PL spectra at zero magnetic field. We observe both the lower polariton (LP) and upper polariton (UP) dispersion branches with a characteristic avoided crossing of the two emissions.

The subsequent panels in Fig.~\ref{f1} illustrate the PL spectra in the magnetic field at  5~T, 7.5~T and 14~T. We observe that the magnetic field modifies the emission spectrum in three main aspects: the energy shift, polariton population and linewidth. Below, we discuss in detail the observed effects.

\section{Magnetopolariton energy}
\label{sec:energy}

Let us first focus on the change of the polariton emission energy in the magnetic field. In Fig.~\ref{f1} we directly observe a global blue-shift of polariton states with increasing field strength, with a magnitude that strongly depends on the polariton wavevector. The energy shift of LP state at $\theta$~=~0~ deg. ($k_{\parallel}$~=~0~$\mu m^{-1}$) is smaller than at high angles (wavevectors) resulting in a deep LP dispersion at high fields. The effect caused by the change of LP and UP emission energies under magnetic field is similar to the effect of the change of the detuning towards more negative values.

This is expected because the magnetic field is acting mainly on the excitonic component of polaritons. The photonic part stays essentially unchanged and experimental evidence is given further on in this section. 

A polariton dispersion shape can be described by the model of two coupled oscillators \cite{weisbuch}. The coupling Hamitonian of the two-level system in the basis of the uncoupled oscillators is given by:
\begin{equation}
H=
\begin{pmatrix} 
E_{x}(B) & \hbar\Omega/2 \\ 
\hbar\Omega/2 & E_{c}(k_{\parallel})
\end{pmatrix}
\label{eq0}
\end{equation}
where $\Omega$ is the coupling strength. The dispersions of the LP ($E_-$) and UP ($E_+$) polariton branches are therefore
\begin{equation}
E_{\pm}(k_{\parallel},B)=\frac{E_{c}(k_{\parallel})+E_{x}(B)}{2} \pm \frac{1}{2} \sqrt{\delta^{2}(k_{\parallel})+\hbar^{2}\Omega^{2}(B)},
\label{eq2}
\end{equation}
where $\delta=E_{c}(k_{\parallel})-E_{x}(B)$, $E_{c}(k_{\parallel})$ describes the photonic dispersion due to the confinement of photonic mode in the microcavity
\begin{equation}
E_{c}(k_{\parallel})= \frac{\hbar c}{n}\sqrt{k_{z}^{2}+k_{\parallel}^{2}},
\label{eq3}
\end{equation}
and $E_{x}(B)$ is the magnetoexciton energy and $n$ is the (average) cavity refractive index. We neglect here the exciton Zeeman splitting. We present the results of the polarisation-resolved experiments and discuss the Zeeman effects in Sec.~\ref{sec:zeeman}. We also neglect the exciton dispersion that has no effect on the very narrow momentum space region studied here.

Dispersion curves, plotted in Fig.~\ref{f1}, are the result of the fit of Eq.~\ref{eq2} to the experimental data. The fitting parameters are the exciton energy, $E_{x}(B)$, the photon energy at zero in plane momentum, $E_{c}(k_{\parallel}=0)$, and the exciton-photon coupling energy, $\hbar\Omega$. The variation of $\Omega$ under magnetic field is discussed in Sec.~\ref{sec:Rabi}. The changes of exciton and photon energies and the resulting detuning with magnetic field are illustrated in Fig.~\ref{f2}. Indeed, we observe that the photon energy does not change with magnetic field and the exciton energy increases. The magnetic field changes the electron and hole wave functions, leading to both (i) an exciton energy shift and (ii) an increase of the exciton oscillator strength. The increase of the exciton emission energy is due to the influence of magnetic field on the electron-hole relative motion, resulting in the diamagnetic shift of the exciton energy and quantisation of the conduction and valence bands into a series of Landau levels, that is significant at high magnetic fields.

The blue-shift of the exciton energy in magnetic field observed in our experiment agrees well with the typical magnetoexciton behaviour in QWs. This topic has been widely studied theoretically and experimentally in both low- and high- field regimes. Here we compare our results with the model presented in Ref.~\cite{MacDonald} assuming that our excitons in narrow QW are similar to the two-dimensional hydrogen atom, as well as with a model that takes into account the dimension perpendicular to the QW~\cite{piotrek-Bexciton}. A good theoretical fit of~\cite{MacDonald} is obtained for the exciton energy shift in magnetic field at any field strength, as illustrated in Fig.~\ref{f2} a). Based on this model we could determine the exciton binding energy to be approx. 7~meV and the exciton effective mass 0.046~$m_0$. Both are reasonable values taking into account the mixed alloy composition in our InGaAs QW. 

The model, where the exciton energy and electron-hole wave functions are calculated self-consistently by a variational method,
and the third dimension of the finite quantum well is taken into account~\cite{piotrek-Bexciton},
provides an analytical expression for $E_x(B)$ that is reliable in a relatively large region of low magnetic fields, (see the solid line in Fig.~\ref{f2}-a). 
The agreement between the two theoretical methods and the experimental results is excellent in this region.

Focusing now on the detuning, $\delta(B)= E_{c}-E_{x}(B)$, it becomes more negative as the exciton energy $E_{x}$ increases with the field (Fig.~\ref{f2} b)). In other words, the magnetic field induces a negative detuning. The detuning changes by 8~meV from 0~T to 14~T. It becomes linear above 8~T with a slope of 0.67~meV/T.

Equation~(\ref{eq2}) has another important consequence. Along the dispersion the excitonic and photonic contents change, as described by the Hopfield coefficients \cite{hopfield}. The polaritons on the LP branch with zero wavevector are more photonic, and those with high wave vectors are mostly excitonic. The more excitonic the polariton state, the higher the energy shift observed in magnetic field. We present cross-sections of the polariton emission from Fig.~\ref{f1} a) for two different emission angles: ($\theta$~=~0~deg.) in Fig.~\ref{f3} a), and a large angle, ($\theta$ = 20 deg.), in Fig.~\ref{f3} b), in magnetic fields up to 14T. We observe a non-monotonic behaviour of both the LP and the UP branch. The value of the energy shift scales directly with the excitonic content in the polariton state \cite{armitage, forchel}.

The values of the energy, the emission intensity and the linewidth of both polariton lines can be obtained directly from the experiment. For each angularly resolved PL image in the magnetic field (as presented in Fig.~\ref{f1}) and for each emission angle we fitted a lorentzian function describing the line-shape of the polariton emission.  From such a fit we retrieve directly the polariton energy, intensity and linewidth. Fig.~\ref{f4} illustrates the change of the emission energy of LP and UP under magnetic field for all emission angles. The higher the emission angle, the faster the energy change for LP. The effect is reversed for UP.

\section{Exciton oscillator strength}
\label{sec:Rabi}

The change of the exciton oscillator strength follows from an increased electron and hole wave functions overlap imposed by the magnetic field.  The increase of the exciton oscillator $f_{osc}$ strength is directly visible in our data by the change of the Rabi splitting $\Omega$ according to \cite{savona}
\begin{equation}
\Omega\sim\sqrt{\frac{N f_{osc}}{L_{eff}}},
\label{eq4}
\end{equation}
where $N$ - number of quantum wells, and $L_{eff}$ - effective cavity length, are constant and depend on the sample structure. The above relation is valid if the exciton and cavity modes have the linewidths significantly narrower than the Rabi splitting, which is the case here.

At negative detuning, Rabi splitting is given by the minimum of energy separation between the UP and LP branches. In order to obtain the value of Rabi splitting directly from the experiment, we calculated the difference between the UP and LP energy (maps in Fig.~\ref{f3}) for each magnetic field. Fig.~\ref{f5} illustrates this difference. The minimum value on the map corresponds to the Rabi splitting. It can be traced along the dashed lines plotted in Fig.~\ref{f5}. The anticrossing position corresponds to a different emission angle due to a change of the detuning induced by the magnetic field. The Rabi splitting can also be obtained following the fit of Eq.(~\ref{eq2}) to the data presented in Fig.~\ref{f3}. The result is illustrated in Fig.~\ref{f6}. We observe a monotonic increase of $\Omega$ from approx.~3.4meV to 4.8~meV at 14~T, in very good agreement with Fig.~\ref{f5}. 

According to Eq.~(\ref{eq4}), we have a direct access to the exciton oscillator strength $f_{osc}$. The normalised value of $f_{osc}$ defined as: $f_{norm}~=~f_{osc}(B)/f_{osc}(0)$ is illustrated in Fig.~\ref{f6} b). We observe that $f_{norm}$ increases be a factor of two up to 14~T. At low field strength $f_{norm}$ is almost constant as the Coulomb energy dominates over cyclotron energy and the exciton wavefunction is not perturbed. At higher field strengths, the magnetic field quantisation is strong enough to shrink the exciton wavefunction and $f_{norm}$ starts to increase rapidly. This result is in perfect agreement with previously experimentally observed and numerically calculated enhancement of the exciton oscillator strength under magnetic field in various types of the microcavity structures \cite{whittaker, fisher, berger, tanaka, armitage} and with the calculations performed for the structure investigated here presented in Ref.~\cite{zieba}.

Let us now comment on the small wiggle observed in our data between 1.5~-~1.8~T, well visible in Fig.~\ref{f6}. This effect is caused by some additional resonance which disturbs the shape of the polariton branches. This resonance is due to the coupling of the cavity photon to the excited exciton state, 2$s$. This resonance occurs at high wavevectors and is not visible directly in the experimental data presented here, but affects the estimation of the energy of the 1$s$-exciton-polariton. 

\section{Magnetopolariton Zeeman splitting}
\label{sec:zeeman}

Due to the Zeeman splitting of exciton states, the magnetic field imposes polarisation of polariton states, manifested as a change in the polarisation of the detected light. In our model the two spin components of excitons become decoupled and therefore couple independently to photons of the corresponding circular polarisation. To introduce the excitonic Zeeman effect in polariton states, we rewrite Eq.~(\ref{eq2}) for each spin component. In the following, we will restrict our considerations to the LP polariton branch (the calculations for the UP case are analogous).
Replacing  $E_x(B)$ by $E_{x,\sigma^\pm}(B)$ we get
\begin{equation}
E_{x,\sigma^\pm}(B)=E_x(B) \pm \frac{1}{2} g \mu_B B,
\label{eq5}
\end{equation}
where from now on $\pm$ corresponds to the spin-up and spin-down components, in contrast to  Eq.~(\ref{eq2}).
We obtain the LP branch energy
\begin{align}
E_{LP,\sigma^\pm }(B)&=\frac{E_{c}+E_{x,\sigma^\pm}(B)}{2} -\\\nonumber 
&- \frac{1}{2} \sqrt{\left(E_{c}-E_{x,\sigma^\pm}(B) \right)^{2}+\hbar^{2}\Omega^{2}(B)}.
\label{eq5b}
\end{align}
The Zeeman splitting of the LP is given by
\begin{equation}
E_{Z, LP}(B)=E_{LP,\sigma^+}(B) - E_{LP,\sigma^-}(B),
\label{eq6}
\end{equation}
which gives a simple expression
\begin{equation}
E_{Z, LP}(B)= \frac{1}{2}g \mu_B B + \frac{1}{2} \sqrt{\delta_-^2+(\hbar \Omega)^2} -  \frac{1}{2} \sqrt{\delta_+^2+(\hbar \Omega)^2},
\label{eq7}
\end{equation}
where
\begin{equation}
\delta_\pm = E_c - E_x(B) \mp \frac{1}{2} g \mu_B B
\label{eq8}
\end{equation}
is the effective detuning of the spin-up and spin-down components.

In the limit of low magnetic field and small detuning, $|\delta_\pm|\ll\hbar\Omega$, it is possible to obtain a simpler form of the LP Zeeman splitting using the 
alternative expression for the eigenvalues of the Hamiltonian~(\ref{eq0})
\begin{equation}
E_{LP,\sigma^\pm} = \mathcal{X}_{\sigma^\pm}^2 E_{x,\sigma^\pm} + \mathcal{C}_{\sigma^\pm}^2 E_{c} - \mathcal{X}_{\sigma^\pm} \mathcal{C}_{\sigma^\pm} \hbar\Omega,
\end{equation}
where $\mathcal{X}_{\sigma^\pm}, \mathcal{C}_{\sigma^\pm}$ are the polariton Hopfield coefficients (determining the excitonic and photonic content)~\cite{hopfield}, with
$\mathcal{X}^2_{\sigma^\pm} = (1+(1+(\hbar\Omega/\delta_\pm)^2)^{-1/2})/2$ and $\mathcal{X}_{\sigma^\pm}^2+ \mathcal{C}_{\sigma^\pm}^2=1$. The LP Zeeman splitting can be now written as
\begin{equation}
E_{z,LP} =  \mathcal{X}^2 g\mu_B B + O(|\delta_\pm|/\hbar \Omega),
\end{equation}
i.e. the polariton Zeeman splitting is approximately equal to the bare exciton splitting times the excitonic content. This formula is particularly useful in the
low magnetic field limit, but fails in the strong field regime, where the detunings $\delta_\pm$ become large and negative.

In the regime where the detunings 
are large with respect to the Rabi splitting, $|\delta_\pm|/\hbar\Omega \gg 1$, we can expand the left hand side of Eq.~(\ref{eq7}) in Taylor series to obtain
\begin{equation}
E_{Z,LP} \approx g\mu_B B\frac{\hbar^2\Omega^2}{4\delta_+\delta_-},
\end{equation}
In the case of the strongest fields attained in the experiment (10-14 T), the theoretical and experimental data is well described with
simple formulas $\delta_\pm\approx C_0+(C_1\pm \frac{1}{2} g \mu_B) B$, and $\Omega \approx \Omega_0$, where $C_0, C_1$ and $\Omega_0$ are constants. 
Interestingly, the LP Zeeman splitting decreases with the magnetic field strength in the strong field regime.

Experimentally, we resolve the polariton emission in circular polarisations, $\sigma_{+}$ and $\sigma_{-}$. Fig.~\ref{f1-1} illustrates the experimental dispersion curves for one polarisation ($\sigma+$). The data in $\sigma-$ polarisation are the same in this scale as the Zeeman splitting is very small. The data in Fig.~\ref{f1-1} are purposely taken for positive detuning on the sample, $\delta_{0}=$+1.4~meV, because excitonic effect are more pronounced, even at strong field strengths.

Further on, we perform the same data analysis with a lorentzian fit to all spectra at a given wavevector and for all magnetic field values, from which we retrieve the emission energy for LP and UP for both polarisations. The energy difference for LP branch between $\sigma_{+}$ and $\sigma_{-}$ emissions is illustrated in Fig.~\ref{f9} a) and can be compared directly with our calculations based on Eq.~\ref{eq7} in Fig.~\ref{f9}~b). 

We observe that the value of polariton Zeeman splitting depends directly on the excitonic content in polariton state. The more the state is excitonic, the larger is the splitting with the highest value for pure exciton state. In particular, the polariton Zeeman splitting at $\theta$=~0~deg. for LP can be smaller at 14~T (80$\mu eV$) than at  5.5~T (175~$\mu eV$). 
As the results plotted here are for positive exciton-photon detuning at B~=~0~T, the LP is mostly excitonic-type along the whole dispersion up to 5~T. At 5~T we retrieve zero detuning case with equal exciton and photon component in polariton state. At higher field strengths, the photonic component in LP branch start to dominate for low emission angles. For $\theta$~=~0~deg., above 5~T we observe the decrease of the polariton Zeeman splitting, in accordance with our theoretical analysis. 

In our theoretical calculations (Fig.~\ref{f9}~b) we use the values of $E_x(B)$ evaluated in Sec.~\ref{sec:energy} with the $g$-factor as the only fitting parameter. The cross sections of the LP Zeeman splitting maps from Fig.~\ref{f9} at two emission angles are illustrated in Fig.~\ref{f10}. We obtained the best fit for $g~=~1.35$, which is in a good agreement with the exciton's $g$-factor in 8-nm-thick Ga$_{1-x}$In$_x$As QW with the In content of $x$~=~0.04 reported in the literature \cite{wang, traynor}.

\section{Magnetopolariton intensity and linewidth} 
\label{sec:intensity}

The photoluminescence experiment gives us the access not only to the emission energy but also to the intensity and linewidth of the emitted light. The intensity is directly proportional to the number of polaritons occupying a state of given energy and in plane momentum \cite{savona}. The distribution of polariton population along the dispersion curve is a result of an interplay between the relaxation, polariton-acoustic phonon interaction and thermalisation. In Fig.~\ref{f7} and Fig.~\ref{f8} we present the results obtained directly from the lorentzian fit to the experimental data of the polariton intensity and linewidth, respectively. 

Upon increasing magnetic field we observe both the total PL intensity decrease and the accumulation of polaritons at the bottleneck at LP branch. The bottleneck effect is expected as the magnetic field induces negative detuning. The detuning $\delta_0$, defined by Eq.~\ref{eq00}, for each magnetic field strength is marked directly in the figures on the top-axis. At high field strengths the population at the bottleneck decreases due to the reduced scattering between excitons and acoustic phonons as the polaritons becomes more photon-like. Apart from the high intensity at the bottleneck region (i.e. at high emission angles) we observe surprisingly strong population of polaritons at the bottom of LP branch at $\theta$~=~0~deg. The intensity at the bottom of LP branch and at the bottleneck region are compared in Fig.~\ref{f11}. The intensity at the bottom of LP branch dominates over the intensity at the bottleneck. 

This effect suggests an increased scattering process induced by the magnetic field that can transform polaritons directly to the LP ground state. In fact, the magnetic field shifts the excitonic levels in the quantum well, but also the energy of the QW barrier material, the GaAs. The energy difference between the bottom of LP branch and the barrier excitons with increasing magnetic filed approaches 36~meV, where it matches exactly the LO phonon in GaAs. The tail of the Boltzmann distribution of exciton population in GaAs can be therefore captured through optical phonons emission. The relaxation process with optical phonons is hundred times more effective than with acoustic phonons. Therefore, in magnetic field we observe both effects: high population at LP ground state (due to the increasing LO-phonon scattering) and at the bottleneck (due todecreasing LA-phonon scattering).

At very high field strengths we expect also the modification of the exciton dispersion in the region of high wavevectors, above 100~$\mu m^{-1}$ (not accessible in our experiment). The exciton dispersion becomes flat at high wavevectors \cite{chemla}. The magnetic field therefore strongly modifies also the excitonic reservoir, where the exciton effective mass increases.

The variations of LP and UP polariton linewidths in function of magnetic field are illustrated in Fig.~\ref{f8}. The data presented here are only for the $\sigma^+$ polarisation, in order to eliminate the broadening effects due to the Zeeman splitting (very similar results are obtained for the  $\sigma^-$ polarisation). At low magnetic field strength and at low emission angles the predominantly excitonic LP branch has a linewidth of approx. 0.1~meV and 
the value decreases to 0.087~meV at 14~T, when the states becomes cavity like. This situation is reversed for the UP branch. At low magnetic field and low emission angles the line is narrow (approx. 0.13~meV) as the state is mostly photonic. The line broadens to 0.8~meV at 14~T, when the state is mostly excitonic. The zero-detuning angle in each magnetic field is marked directly in Fig.~\ref{f8} with the white dashed line and corresponds to the resonant situation where the linewidth of LP and UP are expected to be equal. We observe smaller emission linewidth for the LP polariton, which was often observed previously \cite{houdre}. 

Fig.\ref{f12} shows the cross-sections of the map illustrating the linewidth of LP and UP (Fig.~\ref{f8}) along the minimum energy of polariton branches (corresponding to zero emission angle) and along the zero detuning angle (white dashed lines in Fig.~\ref{f8}). The zero detuning angle represents the resonant situation where the LP and UP are equally exciton and photon like. At each magnetic field its value is different due to the change of the global detuning $\delta_0$. 
The observed general tendency is the increase of polariton linewidth in magnetic field, even in the anticrossing points where both branches share the same excitonic and photonic components (red curves in Fig.~\ref{f11}).  The increase of polariton linewidth is even more pronounced for UP traced along zero emission angle ($k_{II}=0$, black-dashed curve in Fig.~\ref{f11}), as this state becomes more excitonic-like upon increasing field strength. The magnetic field imposes an additional in-plane confining potential to excitons and shrinks the electron and hole wave functions. 

As we directly probe the linewidth in $k$-space, the field-induced linewidth broadening corresponds exactly to the wavefunction shrinkage in real space. The exciton relative motion wavefunction, responsible for the polariton linewidth \cite{ell}, becomes more sensitive to the disorder potential and the number of available collision states increases, which was also observed in Ref. \cite{berger}. This is a reversed process to a motional narrowing effect \cite{whittaker_motional, armitage}, or equivalently, motional narrowing process becomes less pronounced. As the field increase, the statistical averaging over the disorder potential is less effective and the polariton line broadens. The linewidths of polariton states that become more and more photon-like (black solid curve in Fig~\ref{f11}) slightly decrease in magnetic field as the photonic function is not perturbed by the magnetic field.
Such a situation is not observed when the cavity photon is in resonance with high energy exciton-continuum states. In magnetic field the continuum states are quantised in Landau levels and the cavity linewidth narrowing is observed due to the transition form the two- to zero-dimensional electronic density of states \cite{tignon, harel}.

\section{Summary}

We studied the effect of perpendicularly applied external magnetic field on the photoluminescence spectra of exciton-polaritons in a semiconductor microcavity. The design of the optical setup allowed us to probe the angularly resolved emission giving access to the full polariton dispersion (energy - $k$-space dependence). The change of lower and upper polariton dispersion was observed in magnetic field up to 14~T. At each magnetic field strength we were able to trace the polaritons at zero exciton-photon detuning with no change of the position on the sample, temperature or any other external parameter. 
We have shown a strong modification of the polariton energy, population and linewidth induced by the external magnetic field. The excitonic resonance can be tuned up to 8~meV across the photon mode and the exciton oscillator strength increases by a factor of two up to 14T, which directly influences the increase in the Rabi splitting. The observed Zeeman splitting of polariton states can be very well modelled by taking into account the independent coupling of two exciton spin components with cavity photon polarisations. The modification of polariton intensity and linewidth are well described by taking into account the influence of external magnetic field on the exciton dispersion and in-plane relative motion, while the photonic resonance is not influenced by the magnetic field.

\section{Acknowledgements}

B.P., J.S., J. \L{}, P. Z. and I. T. acknowledge the support from the National Science Center (NCN) grant 2011/01/D/ST7/04088. The part of the experimental work was possible thanks to the Physics facing challenges of XXI century grant at the Warsaw University, Poland. P. S. and M. M. acknowledge the support from the NCN grant 2011/01/D/ST3/00482. M.R.M. kindly acknowledge the NCN (DEC-2013/08/T/ST3/00665 and DEC-2013/09/N/ST3/04237) for financial support for his Ph.D. Partial support from the European Research Council (ERC AG  ''MOMB'', No. 320590) is also acknowledged.




\end{document}